\documentclass[final]{cvpr}
\usepackage{times}
\usepackage{epsfig}
\usepackage{graphicx}
\usepackage{amsmath}
\usepackage{amssymb}
\usepackage{comment}
\usepackage{makecell}
\usepackage{soul}
\usepackage{multirow}
\usepackage{enumerate}
\usepackage[pagebackref=true,breaklinks=true,colorlinks,bookmarks=false]{hyperref}

\def\confYear{MICCAI FLARE 2021}

\begin{document}

%%%%%%%%% TITLE
\title{Efficient Context-Aware Network for Abdominal Multi-organ Segmentation}

\author{Fan Zhang\textsuperscript{1}, Yu Wang\textsuperscript{2}, Hua Yang\textsuperscript{3}\\
\textsuperscript{1, 2, 3}Department of Radiological Algorithm, Fosun Aitrox \\
Information Technology Co., LTD., Shanghai, China.\\
\tt\small zhangfan2@fosun.com\textsuperscript{1}, wangyu@fosun.com\textsuperscript{2}, yanghua@fosun.com\textsuperscript{3}
}

\maketitle

\begin{abstract}
The contextual information, presented in abdominal CT scan, is relative consistent. In order to make full use of the overall 3D context, we develop a whole-volume-based coarse-to-fine framework for efficient and effective abdominal multi-organ segmentation. We propose a new \textbf{efficientSegNet} network, which is composed of basic encoder, slim decoder and efficient context block. For the decoder module, anisotropic convolution with a k*k*1 intra-slice convolution and a 1*1*k inter-slice convolution, is designed to reduce the computation burden. For the context block, we propose strip pooling module to capture anisotropic and long-range contextual information, which exists in abdominal scene. Quantitative evaluation on the FLARE2021 validation cases, this method achieves the average dice similarity coefficient (DSC) of 0.895 and average normalized surface distance (NSD) of 0.775. This method won the 1st place on the \href{https://flare.grand-challenge.org/FLARE21/}{2021-MICCAI-FLARE} challenge. Codes and models are available at \href{https://github.com/Shanghai-Aitrox-Technology/EfficientSegmentation}{https://github.com/Shanghai-Aitrox-Technology/EfficientSegmentation}
\end{abstract}

\section{Introduction}
\begin{figure}[htbp]
\centering
\includegraphics[scale=0.28]{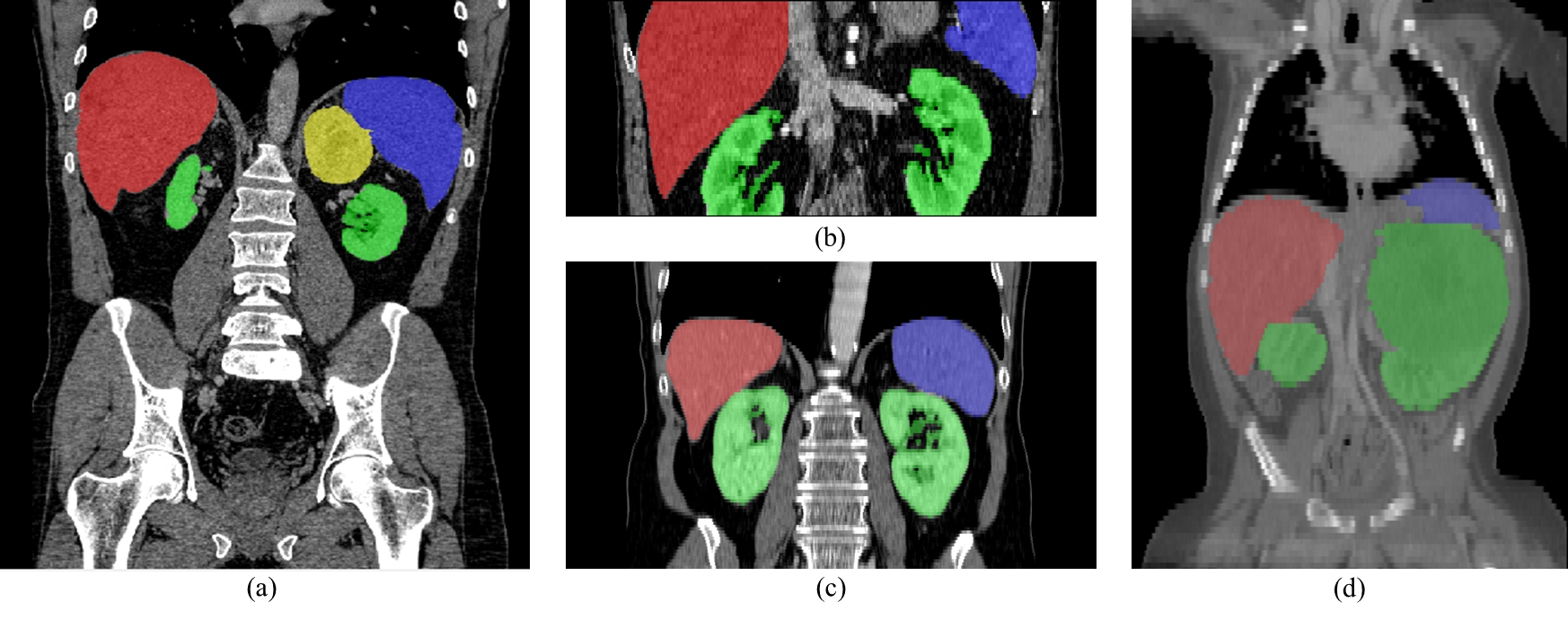}
\caption{An illustration of abdominal CT images vary in field-of-views (a-c), and lesion-affected organ (d) on FLARE2021 dataset.}
\label{fig:View}
\end{figure}

In this paper, we focus on multi-organ segmentation from abdominal CT scans. As shown in Figure~\ref{fig:View}, the main difficulties stem from four aspects: 
1)  The variations in field-of-views, shape and size of different organs. 
2)  The abnormalities, like lesion-affected organ, may lead to segmentation failure.
3)  The diversity of data source in term of multi-center, multi-phase and multi-vendor cases.
4)  The limited GPU memory size and high computation cost.\\

A common solution~\cite{isensee2021nnu} is to develop a sliding-window method, which can balance the GPU memory usage. Usually, this method need to sample sub-volumes overlap with each other to improve the segmentation accuracy, while leading to more computation cost. Meanwhile, sub-volumes sampled from entire CT volume inevitably lose some 3D context, which is important for distinguishing multi-organ with respect to background.\\

We develop a whole-volume-based coarse-to-fine framework~\cite{coarse_to_fine} to effectively and efficiently tackle these challenges. The coarse model aims to obtain the rough location of target organ from the whole CT volume. Then, the fine model refines the segmentation based on the coarse result. This coarse-to-fine pipeline can cover anatomical variations for different cases. To capture the spatial relationships between multi-organ, we exploit strip pooling~\cite{stripPooling} for collecting anisotropic and long-range context. This strip pool offers two advantages. Firstly, compared to self-attention or non-local module, strip pool consumes less memory and matrix computation. Secondly, it deploys long but narrow pooling kernels along one spatial dimension to simultaneously aggregate both global and local context.\\

The main contributions of this work are summarized as follows:
\begin{enumerate}[1)]
\item We propose a whole-volume-based coarse-to-fine framework to make full use of the overall 3D context, this pipeline covers the anatomical variations occurred on different cases.
\item We design anisotropic convolution block with low computation cost. We propose strip pooling module to capture anisotropic and long-range contextual information.
\item The effectiveness and efficiency of the proposed whole-volume-based coarse-to-fine framework are demonstrated on FLARE2021 challenge dataset, where we achieve the state-of-the-art with low time cost and less memory usage.
\end{enumerate}

%###########################
%###########################
\begin{figure*}[htb]
\centering
\includegraphics[scale=0.35]{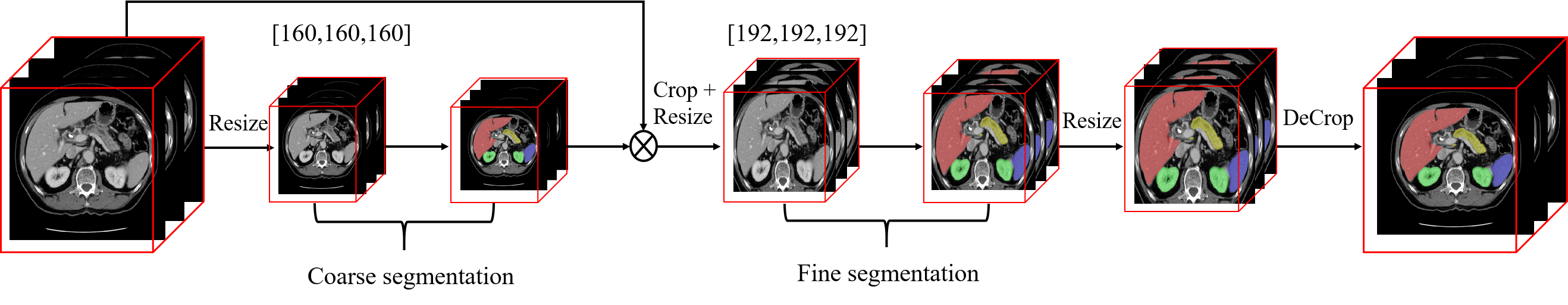}
\caption{A schematic diagram of whole-volume-based coarse-to-fine segmentation framework.}
\label{fig:framework}
\end{figure*}

\section{Method}
As mentioned in Figure~\ref{fig:framework}, this whole-volume-based coarse-to-fine framework is composed of coarse and fine segmentation with a basic U-Net and a carefully designed \textbf{efficientSegNet}, respectively. A detail description of the method is as follows.\\

%###########################
\subsection{Preprocessing}
 The baseline method includes the following preprocessing steps:
\begin{itemize} 
\item Reorienting images to the left-posterior-inferior (LPI) view by flipping and reordering.
\item Resampling image to fixed size. The sizes of coarse and fine input are [160, 160, 160] and [192, 192, 192], respectively.
\item Intensity normalization: First, the image is clipped to the range [-325, 325]. Then a z-score normalization is applied based on the mean and standard deviation of the intensity values.
\end{itemize}
 
\begin{table*}[!htbp]
\caption{Data splits of FLARE2021.}
\label{tab:dataset}
\centering
\begin{tabular}{llll}
\hline
Data Split  
& Center                                                      & Phase                           & \# Num.  \\
\hline
\multirow{2}{*}{Training ( 361 cases )}              & The National Institutes of Health Clinical Center             & portal venous phase           & 80       \\
                                   & Memorial Sloan Kettering Cancer Center                      & portal venous phase             & 281      \\
\hline
\multirow{3}{*}{Validation ( 50 cases )}            & Memorial Sloan Kettering Cancer Center                      & portal venous phase             & 5        \\
                                   & University of Minnesota                                     & late arterial phase             & 25       \\
                                   & 7 Medical Centers                               & various phases                  & 20       \\
\hline
\multirow{4}{*}{Testing ( 100 cases )}              & Memorial Sloan Kettering Cancer Center                      & portal venous phase             & 5        \\
                                   & University of Minnesota                                     & late arterial phase             & 25       \\
                                   & 7 Medical Centers                                & various phases                  & 20       \\
                                   & Nanjing University                                          & various phases                  & 50     \\
\hline
\end{tabular}
\end{table*}

\subsection{Proposed Method}
\begin{figure*}[htbp]
\centering
\includegraphics[scale=0.5]{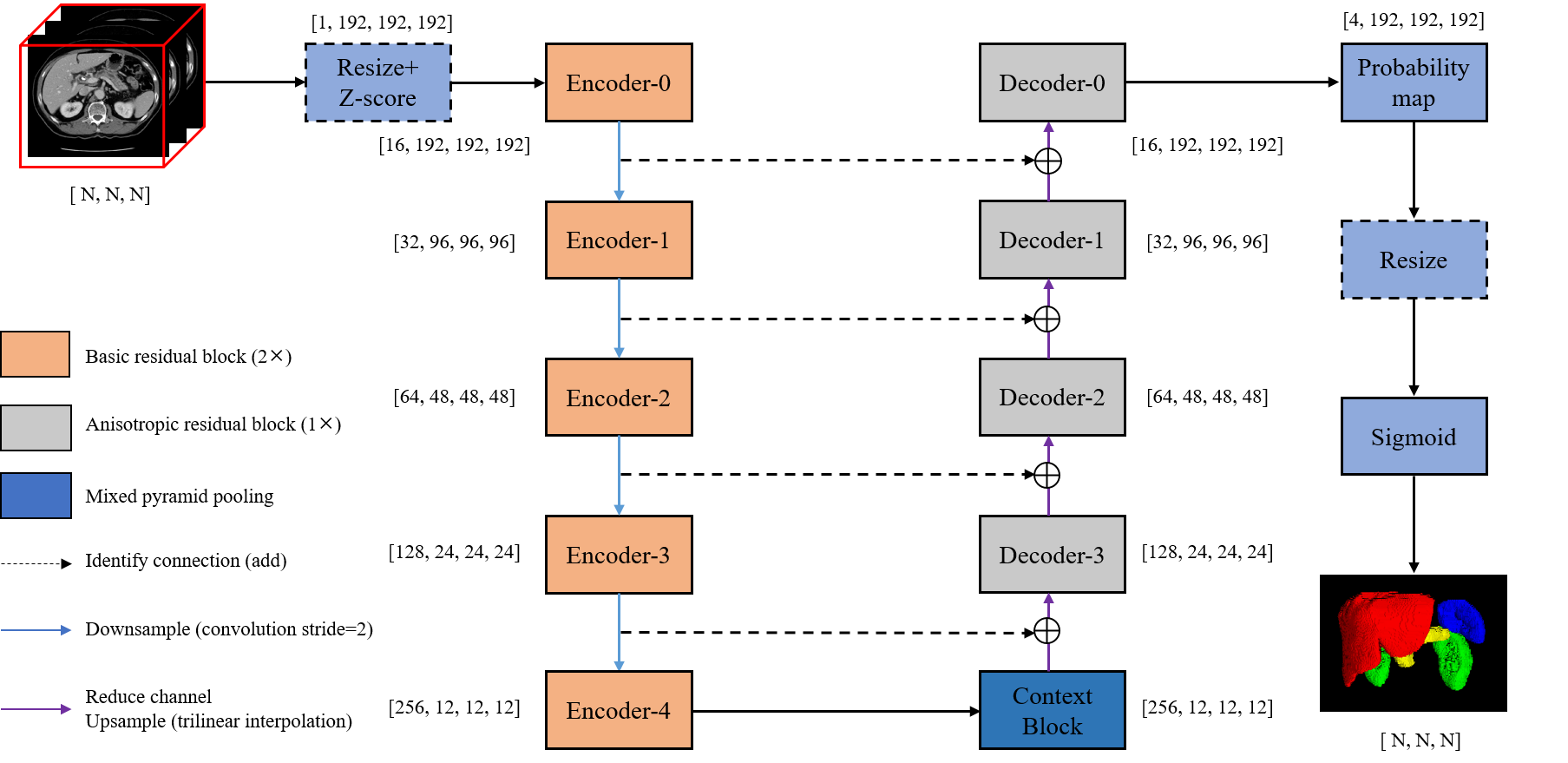}
\caption{Illustration of the proposed efficientSegNet.}
\label{fig:Network}
\end{figure*}

The proposed \textbf{efficientSegNet} consists of three major parts: basic encoder, slim decoder, and efficient context block, as shown in Figure~\ref{fig:Network}.\\

\begin{figure}[htbp]
\centering
\includegraphics[scale=0.5]{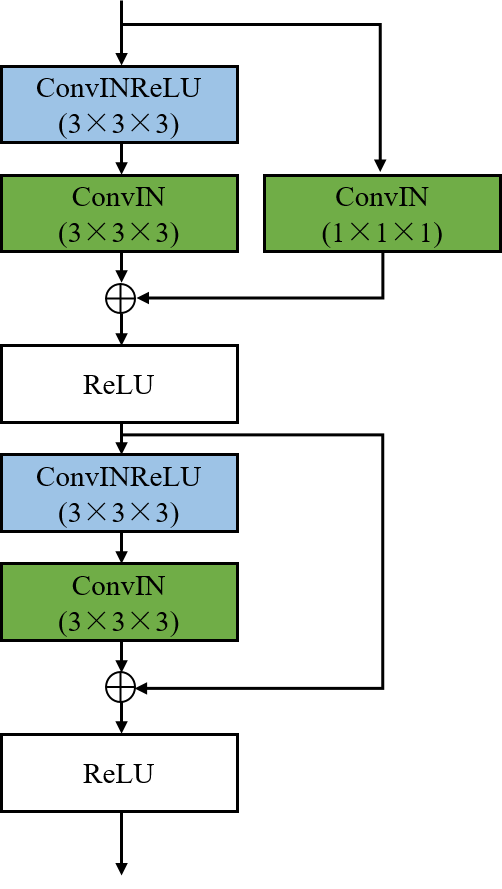}
\caption{Illustration of the encoder block.}
\label{fig:encoder block}
\end{figure}

\begin{figure}[htbp]
\centering
\includegraphics[scale=0.5]{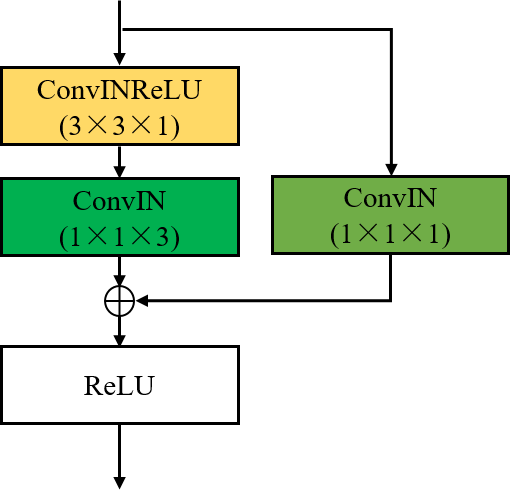}
\caption{Illustration of the decoder block.}
\label{fig:decoder block}
\end{figure}

As depicted in Figure~\ref{fig:encoder block} and Figure~\ref{fig:decoder block}, the encoder module is composed of two residual convolution blocks, and the decoder module with one residual convolution block. As to decoder module, we separate a standard 3D convolution with kernel size 3$\times$3$\times$3 into a 3$\times$3$\times$1 intra-slice convolution and a 1$\times$1$\times$3 inter-slice convolution. The residual convolution block is implemented as follows: conv-instnorm-ReLU-conv-instnorm-ReLU (where the addition of the residual takes place before the last ReLU activation).\\

\begin{figure*}[htbp]
\centering
\includegraphics[scale=0.6]{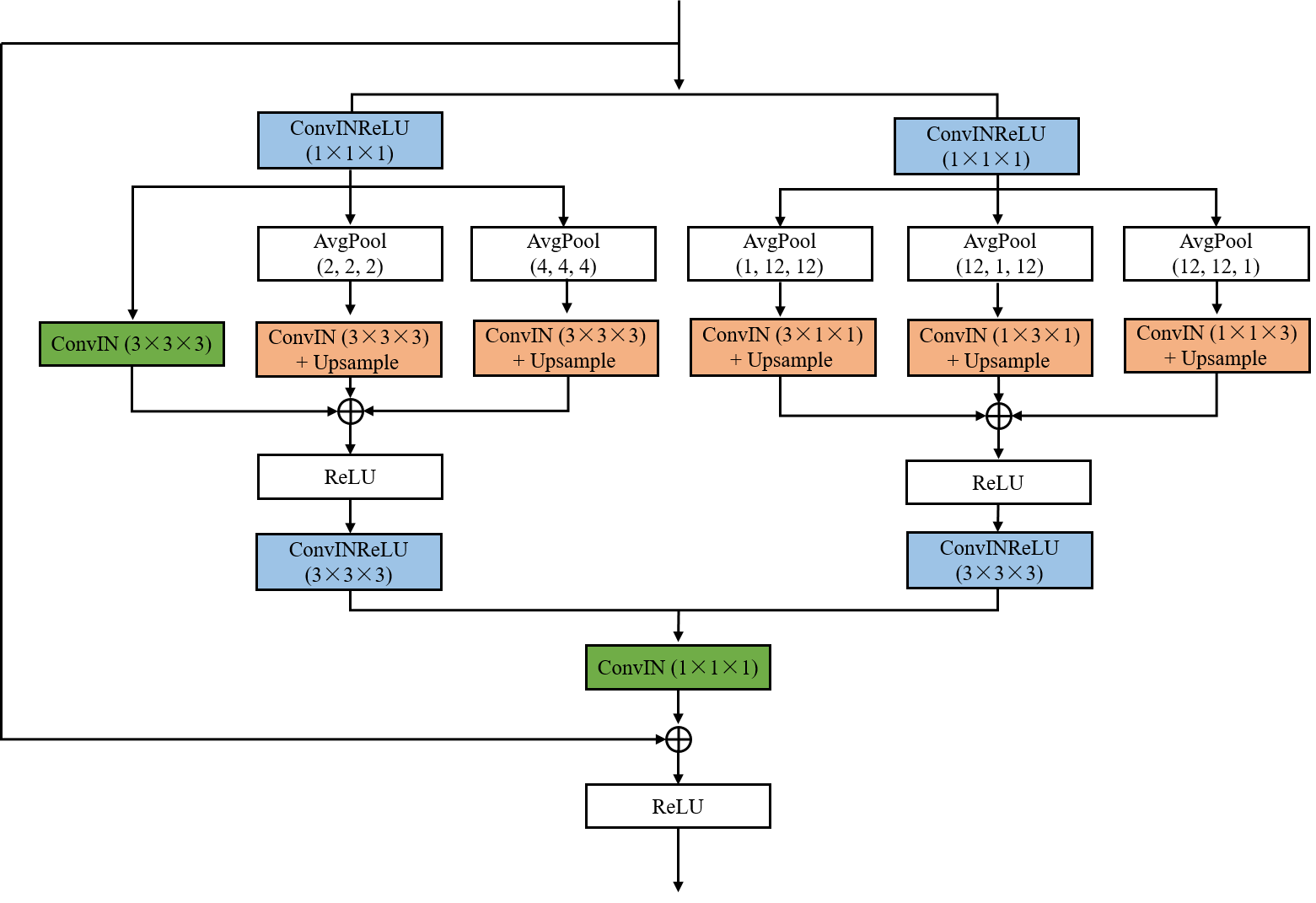}
\caption{Illustration of the context block.}
\label{fig:Illustration of the context block}
\end{figure*}

We adopt 3D-based mixed pyramid pooling (Figure~\ref{fig:Illustration of the context block}) to extract contextual feature, which is composed of the standard spatial pooling and the anisotropic strip pooling. The standard spatial pooling employs two average pooling with the stride of 2$\times$2$\times$2 and 4$\times$4$\times$4. The anisotropic strip pooling with three different-direction receptive fields: 1$\times$N$\times$N, N$\times$1$\times$N and N$\times$N$\times$1, where N is the size of feature map in last encoder module.\\

The initial number of feature maps is 8 for coarse model, while 16 for fine model. We aggregate low and high level feature with addition rather than concatenation, because the former consumes less GPU memory. In addition, the number of model parameters is 9 MB, and the number of flops is 333 GB for 192$\times$192$\times$192 input size.\\

\subsection{Post-processing}
A connected component analysis of segmentation mask is applied on coarse and fine model output.\\

%###########################
%###########################
\section{Dataset and Evaluation Metrics}
\subsection{Dataset}
\begin{itemize}
\item A short description of the dataset used:\\
The dataset used of FLARE2021 is adapted from MSD~\cite{simpson2019MSD} (Liver~\cite{bilic2019lits}, Spleen, Pancreas), NIH Pancreas~\cite{roth9data,roth2015deeporgan,clark2013cancer}, KiTS~\cite{KiTS,KiTSDataset}, and Nanjing University under the license permission. For more detail information of the dataset, please refer to the challenge website and~\cite{AbdomenCT-1K}. The detail information is presented in Table~\ref{tab:dataset}.\\
\item Details of training / validation / testing splits:\\
The total number of cases is 511. An approximate
70\%/10\%/20\% train/validation/testing split is employed resulting in 361 training cases, 50 validation cases, and 100 testing cases. The detail information is presented in Table~\ref{tab:dataset}.
\item Furthermore, the training dataset (361 cases) is randomly divided into training (80\%) and validation (20\%) set, where validation set is used to model selection. A 5-fold cross validation set is generated based on the above mentioned partition.\\
\end{itemize}

\subsection{Evaluation Metrics} 
\begin{itemize}
    \item Dice Similarity Coefficient (DSC)
    \item Normalized Surface Distance (NSD)
    \item Running time
    \item Maximum used GPU memory (when the inference is stable)
\end{itemize}

\begin{table}[!htbp]
\caption{Environments and requirements.}
\label{table:env}
\begin{center}
\resizebox{0.47\textwidth}{!}{
\begin{tabular}{m{2.5cm}<\raggedright|m{6cm}<\raggedright} 
\hline
Ubuntu version       & 16.04.12\\
\hline
CPU   & Intel(R) Xeon(R) Gold 5118 CPU @ 2.30GHz ($\times4$) \\
\hline
RAM                         & 502 GB\\
\hline
GPU                         & Nvidia GeForce 2080Ti ($\times8$)\\
\hline
CUDA version                  & 10.1\\                                                                 
\hline
Programming language                 & Python3.6\\ 
\hline
Deep learning framework & Pytorch (torch 1.5.0, torchvision 0.2.1) \\
\hline
Code is publicly available at                                         &  \href{https://github.com/Shanghai-Aitrox-Technology/EfficientSegmentation}{EfficientSegmentation} \\
\hline
\end{tabular}
}
\end{center}
\end{table}

\begin{table}[!htbp]
\caption{Training protocols.}
\label{table:training}
\begin{center}
\resizebox{0.47\textwidth}{!}{
\begin{tabular}{m{2.5cm}<\raggedright|m{6cm}<\raggedright} 
\hline
\makecell*[c]{Data augmentation\\ methods}                           & Crop and brightness.\\
\hline
\makecell*[c]{Initialization of\\ the network  }                           & Kaiming normal initialization\\
\hline
\begin{tabular}[c]{@{}c@{}} 
Patch sampling \\ strategy \end{tabular}                  & Augment the sample ratio of pathological image (3 times) \\                                                     
\hline
Batch size                    & 16 \\
\hline 
Patch size  & Coarse: 160$\times$160$\times$160 \\ & Fine: 192$\times$192$\times$192 \\
\hline
Total epochs & 200 \\
\hline
Optimizer          & Adam with betas (0.9, 0.99), \\ & L2 penalty: 0.00001\\ 
\hline
Loss          & Dice loss and focal loss \\ & (alpha = 0.5, gamma = 2)\\ 
\hline
Dropout rate          & 0.2\\ 
\hline
Initial learning rate  & 0.01 \\ 
\hline
Learning rate decay schedule & Step decay \\
\hline
Stopping criteria, and optimal model selection criteria & Stopping criterion is reaching the maximum number of epoch (200).\\ 
\hline
Training mode                                           & Mixed precision \\    
\hline
Training time for coarse model    & 3 hours \\
\hline
Training time for fine model      & 6 hours \\
\hline
\end{tabular}
}
\end{center}
\end{table}

%###########################
%###########################
\section{Implementation Details}

\subsection{Environments and requirements}
The environments and requirements of the proposed method is shown in Table~\ref{table:env}.\\

\begin{table*}[htbp]
\normalsize
\centering
%\resizebox{\linewidth}{!}{
\caption{Quantitative results of 5-fold cross validation in terms of DSC and NSD.}
\label{tab:cross-validation}
\renewcommand\tabcolsep{3pt}
\begin{tabular}{cccccccccc}
\hline
\multirow{2}{*}{Training}   & \multicolumn{2}{c}{Liver}  & \multicolumn{2}{c}{Kidney}  & \multicolumn{2}{c}{Spleen}  & \multicolumn{2}{c}{Pancreas}                                       \\
\cline{2-9}   & DSC (\%)   & NSD (\%) & DSC (\%)   & NSD (\%) & DSC (\%)   & NSD (\%) & DSC (\%)   & NSD (\%)  \\
\hline
Fold-1	& 96.8$\pm$5.6	& 88.2$\pm$11.4	& 95.0$\pm$8.4	& 89.6$\pm$12.7	& 95.9$\pm$13.9	& 93.5$\pm$15.1	& 79.3$\pm$23.4	& 68.7$\pm$23.7  \\
\hline
Fold-2	& 96.3$\pm$6.6	& 87.4$\pm$11.3	& 94.0$\pm$11.0	& 88.3$\pm$14.3	& 95.7$\pm$11.5	& 93.2$\pm$13.5	& 79.5$\pm$22.0	& 68.9$\pm$22.2  \\
\hline
Fold-3	& 96.6$\pm$5.1	& 88.1$\pm$10.3	& 94.8$\pm$9.9	& 89.1$\pm$13.0	& 95.6$\pm$13.9	& 93.1$\pm$15.3	& 78.3$\pm$23.0	& 68.0$\pm$23.1  \\
\hline
Fold-4	& 96.3$\pm$6.9	& 88.3$\pm$11.1	& 95.8$\pm$5.4	& 89.5$\pm$12.4	& 95.5$\pm$14.5	& 93.3$\pm$16.1	& 80.9$\pm$21.7	& 69.2$\pm$21.4  \\
\hline
Fold-5	& 96.5$\pm$6.2	& 87.1$\pm$11.7	& 94.5$\pm$10.5	& 88.7$\pm$14.4	& 95.8$\pm$11.6	& 93.5$\pm$13.3	& 79.5$\pm$20.9	& 66.7$\pm$21.7  \\
\hline
Average	& 96.5$\pm$6.1	& 87.8$\pm$11.2	& 94.8$\pm$9.3	& 89.0$\pm$13.4	& 95.7$\pm$13.1	& 93.3$\pm$14.7	& 79.5$\pm$22.2	& 68.3$\pm$22.5  \\
\hline
\end{tabular}
%}
\end{table*}

\subsection{Training protocols}
The training protocols of the proposed method is shown in Table~\ref{table:training}.\\

\subsection{Testing protocols}
The same pre-process and post-process methods are applied as training steps. In order to reduce the time cost of pre-process and post-process, resample and intensity normalization are computed in GPU. We implement the connected component analysis in C++ library, namely cc3d~\cite{CC3D}. We implement the inference model in FP16 mode. Dynamic empty cache is used to reduce GPU memory. \\

%###########################
%###########################
\section{Results}
\subsection{Quantitative results for 5-fold cross validation.}
The provided results analysis is based on the 5-fold cross validation results on training set. Table~\ref{tab:cross-validation} illustrates the results of 5-fold cross validation. While high DSC and NSD scores are obtained for liver, kidney and spleen, DSC and NSD scores for pancreas indicating unsatisfactory performance.\\

\subsection{Quantitative results on validation set.}
The average running time is 9.8 s per case in inference phase, and maximum used GPU memory is 1017 MB. Table~\ref{tab:quanti-validation} illustrates the results on validation set. Compared to the 5-fold cross validation on training set, the results degrade of DSC (approximately 1 point for liver, kidney and spleen, 4 point for Pancreas organ) on validation set is relative low, which indicates highly generation ability. While the results degrade of NSD (approximately 7 point) on validation set is fairly obvious, demonstrating that the boundary regions contain more segmentation errors, which need further improvements.\\

\begin{table}[!htbp]
\caption{Quantitative results of validation set in terms of DSC and NSD.}
\label{tab:quanti-validation}
\centering
\begin{tabular}{ccc}
\hline
Organ    & DSC (\%)        & NSD (\%)        \\
\hline
Liver    & 95.4$\pm$6.60   & 80.4$\pm$13.77  \\
Kidney   & 93.6$\pm$6.50   & 82.8$\pm$12.28  \\
Spleen   & 94.2$\pm$13.88  & 87.1$\pm$16.66  \\
Pancreas & 75.3$\pm$17.44  & 60.5$\pm$16.66  \\
Average  & 89.65           & 77.75            \\
\hline
\end{tabular}
\end{table}

\subsection{Qualitative results}
\begin{figure}[htbp]
\centering
\includegraphics[scale=0.5]{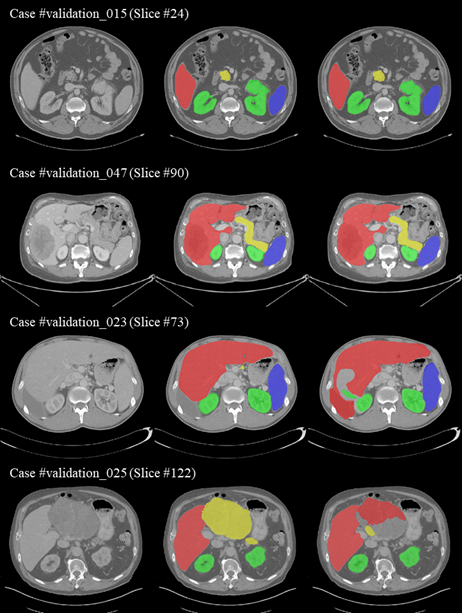}
\caption{Qualitative results on easy (case \#15 and \#47) and hard (case \#23 and \#25) examples. First column is the image, second column is the ground truth, and third column is the predicted results by our propose method.}
\label{fig:results}
\end{figure}

Figure~\ref{fig:results} presents some easy and hard examples on validation set, and quantitative result is illustrated in Table~\ref{tab:examples}. It can be found that the proposed method can segment healthy (case \#15) or slightly lesion-affected (case \#47) organs well, while disappointing performance on seriously lesion-affected (case \#23 and \#25) organs.\\

\begin{table*}[htbp]
\normalsize
\centering
%\resizebox{\linewidth}{!}{
\caption{The DSC and NSD scores of easy and hard examples.}
\label{tab:examples}
\renewcommand\tabcolsep{3pt}
\begin{tabular}{cccccccccc}
\hline
\multirow{2}{*}{Example}   & \multicolumn{2}{c}{Liver}  & \multicolumn{2}{c}{Kidney}  & \multicolumn{2}{c}{Spleen}  & \multicolumn{2}{c}{Pancreas}                                       \\
\cline{2-9}   & DSC (\%)   & NSD (\%) & DSC (\%)   & NSD (\%) & DSC (\%)   & NSD (\%) & DSC (\%)   & NSD (\%)  \\
\hline
Case \#15	& 98.0	& 91.6	& 97.8	& 94.5	& 98.2	& 97.7	& 88.3	& 79.3  \\
\hline
Case \#47	& 98.7	& 93.7	& 95.8	& 87.9	& 97.8	& 94.2	& 92.3	& 80.3  \\
\hline
Case \#23	& \textbf{56.3}	& \textbf{40.4}	& 91.6	& 66.3	& 96.4	& 82.6	& 74.8	& 50.3  \\
\hline
Case \#25	& 84.3	& 71.1	& 96.4	& 87.7	& 97.4	& 95.3	& \textbf{13.5}	& \textbf{14.1}  \\
\hline
\end{tabular}
%}
\end{table*}

%###########################
%###########################
\section{Discussion and Conclusion}

The proposed method can work well on cases where healthy or slightly lesion-affected organs. The proposed method achieves the highly generation ability for liver, kidney and spleen segmentation in terms of DSC scores. Disappointing performance is obtained for pancreas segmentation as a result of the inter-patient anatomical variability of volume and shape. The existence of seriously lesion-affected organ is a critical factor for the poor segmentation performance. Besides, obtaining an accurate boundary segmentation need further investigate.\\

\section*{Acknowledgment}
We sincerely appreciate the organizers with the donation of FLARE2021 dataset. We declare that pre-trained models and additional datasets are not used in this paper.\\

{\small
\bibliographystyle{IEEEtran} 
\bibliography{egbib}
}

\end{document}